%% file: icme2025_template_anonymized.tex
\documentclass[conference]{IEEEtran}
\IEEEoverridecommandlockouts
\usepackage{cite}
\usepackage{amsmath,amssymb,amsfonts}
\usepackage{algorithmic}
\usepackage{graphicx}
\usepackage{array}
\usepackage{booktabs}
\usepackage{textcomp}
\usepackage{xcolor}
\usepackage{siunitx}
\usepackage{url}
\usepackage{float}

\def\BibTeX{{\rm B\kern-.05em{\sc i\kern-.025em b}\kern-.08em
    T\kern-.1667em\lower.7ex\hbox{E}\kern-.125emX}}

\begin{document}

\title{MSDet: Receptive Field Enhanced Multiscale Detection for Tiny Pulmonary Nodule}

\author{Guohui Cai, Ruicheng Zhang, Hongyang He, Zeyu Zhang$^{\dag}$, Daji Ergu, Yuanzhouhan Cao,\\ Jinman Zhao, Binbin Hu, Zhibin Liao, Yang Zhao, Ying Cai$^{*}$
\thanks{Guohui Cai, Daji Ergu, Binbin Hu, and Ying Cai ($^{*}$corresponding author) are with the College of Computer Science and Artificial Intelligence, Southwest Minzu University, Chengdu, China (e-mail: guohuicai123@163.com; ergudaji@163.com; binhu0821@163.com; caiying34@yeah.net).}
\thanks{Ruicheng Zhang is with the School of Intelligent System Engineering, Sun Yat-sen University, Guangzhou, China (e-mail: zhangrch23@mail2.sysu.edu.cn).}
\thanks{Hongyang He is with Department of Computer Science, University of Warwick, UK (e-mail: hongyang.he@warwick.ac.uk).}
\thanks{Zeyu Zhang is with the Australian National University, Canberra ACT 2601, Australia (e-mail: steve.zeyu.zhang@outlook.com). $^{\dag}$Project lead.}
\thanks{Yuanzhouhan Cao is with the School of Computer Science and Technology, Beijing Jiaotong University, Beijing, China (e-mail: yzhcao@bjtu.edu.cn).}
\thanks{Jinman Zhao is with University of Toronto, Toronto, Canada (e-mail:jzhao@cs.toronto.edu).}
\thanks{Zhibin Liao is with the University of Adelaide, Adelaide, Australia (e-mail: zhibin.liao@adelaide.edu.au).}
\thanks{Yang Zhao is with La Trobe University, Melbourne, Australia (e-mail: y.zhao2@latrobe.edu.au).}}


\maketitle

\begin{abstract}
Pulmonary nodules are critical for early lung cancer diagnosis, but traditional CT imaging methods suffer from low detection rates and poor localization. Small nodule detection is challenging due to subtle differences in density and issues like occlusion. Existing methods such as FPN, with its fixed feature fusion and limited receptive field, struggle to effectively overcome these issues. To address these challenges, our paper proposed three key contributions: Firstly, we proposed MSDet, a multiscale attention and receptive field network for detecting tiny pulmonary nodules. Secondly, we proposed the extended receptive domain (ERD) strategy to capture richer contextual information and reduce false positives caused by nodule occlusion. We also proposed the position channel attention mechanism (PCAM) to optimize feature learning and reduce multiscale detection errors, and designed the tiny object detection block (TODB) to enhance the detection of tiny nodules. Experiments on the LUNA16 dataset show an 8.8\% improvement in mAP over YOLOv8, achieving state-of-the-art performance. 
The code is available at \url{https://github.com/CaiGuoHui123/MSDet}.
\end{abstract}

\begin{IEEEkeywords}
Pulmonary nodule; Computer tomography; Tiny object detection; Hybrid CNN-Transformer;
\end{IEEEkeywords}

\vspace{-0.15cm}
\section{Introduction}
\vspace{-0.15cm}
Lung cancer remains the leading cause of cancer-related incidence and mortality worldwide {\cite{siegel2024cancer}}. Pulmonary nodules, often the earliest sign of lung cancer, highlight the importance of early detection. Computed tomography (CT) is the primary method for screening pulmonary nodules. Pulmonary nodules, typically spherical lesions 3–30 mm in diameter, vary in size and morphology, making detection challenging. Small nodules are often difficult to distinguish from surrounding tissues in CT scans, leading to high rates of missed detection and false positives.

\begin{figure}[htbp]
\centering
\includegraphics[width=1\linewidth]{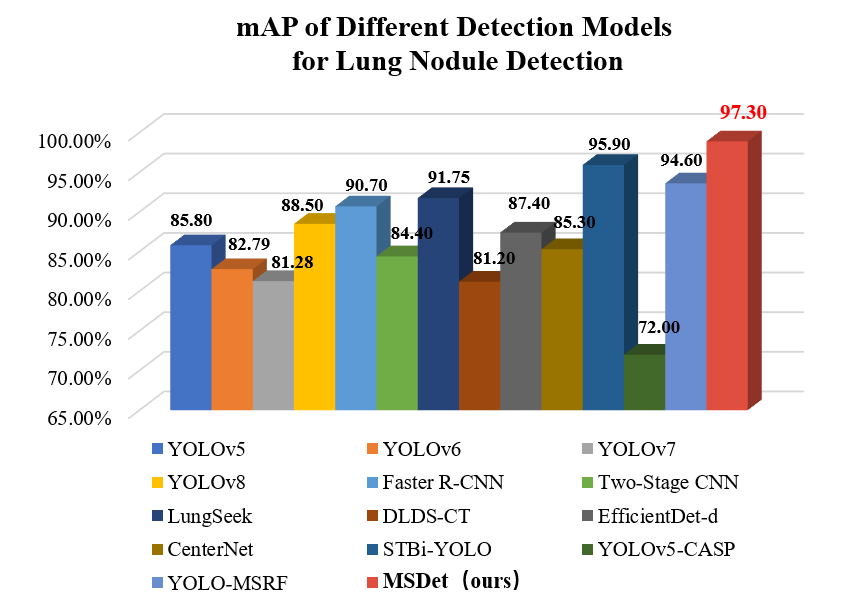} 
\caption{Comparative histogram between the state-of-the-art network and MSDet, \textbf{MSDet (Ours)} achieved the best result of 97.30\% in terms of pulmonary nodule detection accuracy in CT images.}
\label{fig:1}
\end{figure}

Object detection methods are widely used in medical imaging to identify suspicious lesions. Two-stage detection algorithms \cite{girshick2014rich} first detect all potential nodules, often resulting in numerous false positives. These methods are complex and time-consuming, relying on feature maps and prior knowledge (e.g., anchor frames), which can lead to inaccurate localization, especially for small or irregularly shaped nodules. One-stage detection algorithms \cite{redmon2016you} combine candidate detection and false positive reduction, improving speed but struggling with small nodules and occlusions, which may cause missed detections and false positives.

YOLOv5, a popular one-stage detection algorithm, enhances speed and spatial processing but struggles with feature extraction, especially for small targets in complex backgrounds. Transformers \cite{vaswani2017attention}, with their self-attention mechanism, can capture long-range dependencies and global context, improving detection accuracy for small objects like pulmonary nodules by reducing missed detections and false positives.

Inspired by these advancements, we propose MSDet, a novel one-stage model for pulmonary nodule detection, targeting high false positive rates and low accuracy. Our contributions include a tiny object detection block (TODB) to capture finer details for small nodule detection, an extended receptive domain (ERD) strategy to reduce false positives from occlusions, and a positional channel attention mechanism (PCAM) to optimize feature representation. As shown in {Figure \ref*{fig:1}}, MSDet achieved a significant \textbf{8.8\%} improvement in mAP compared to the state-of-the-art nodule detection method YOLOv8.

\begin{figure*}
\centering
\includegraphics[width=1\textwidth]{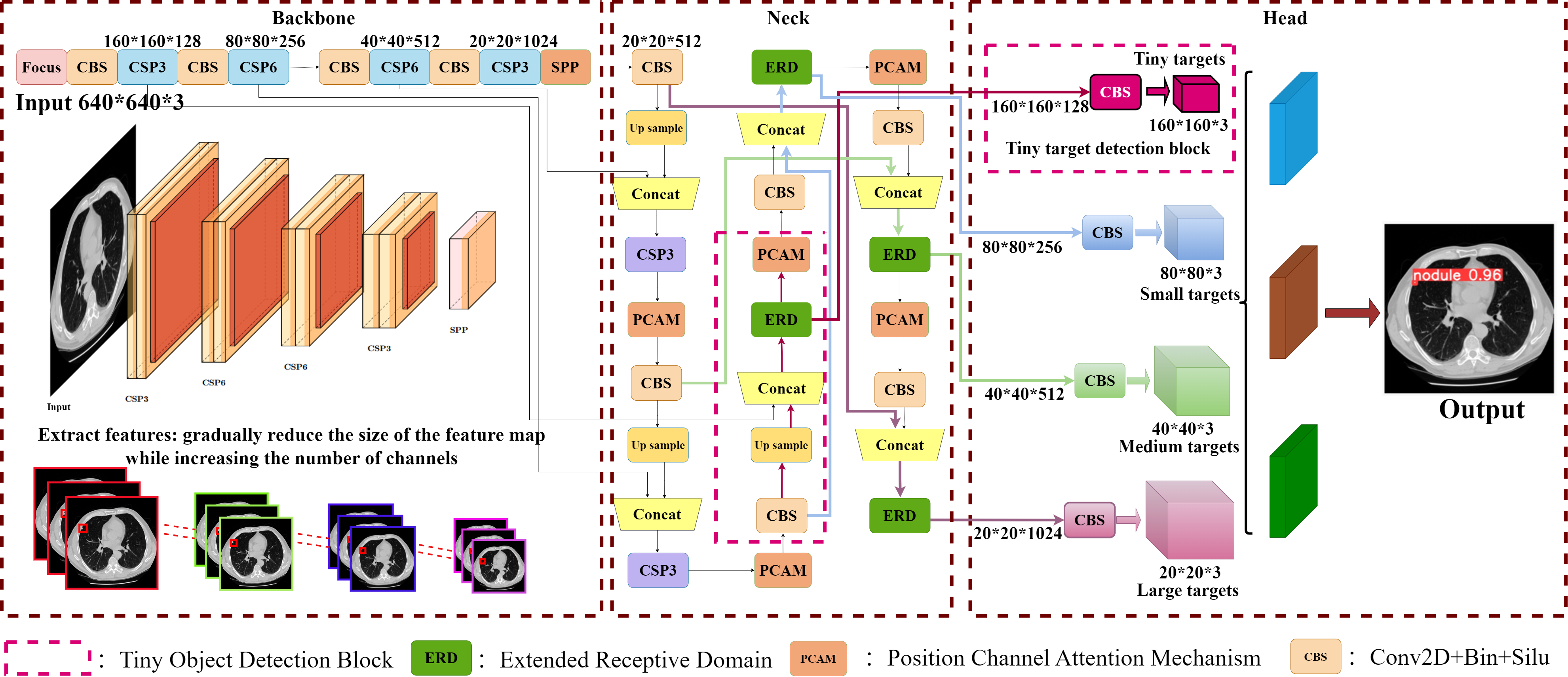} 
\caption{Overall architecture of the MSDet network for lung nodule detection. The initial convolutional layers, represented as CBS blocks, process the input lung CT image to extract preliminary features. These features undergo a series of transformations through the ERD modules, which broaden the receptive field to capture more contextual information. PCAM modules are strategically placed to refine feature representation by focusing on crucial spatial and channel-related information. Multiple feature maps generated at different stages are then concatenated and further processed through upsampling and additional CBS blocks to construct refined prediction feature maps.}
\label{fig:2}
\end{figure*}

Hence, this paper proposed three key contributions as follows:
\begin{itemize}

\item {We proposed \textbf{MSDet}, a novel one-stage detection model specifically designed for detecting tiny pulmonary nodules. MSDet integrates multiscale attention and an enhanced receptive field, addressing challenges of high false positive rates and low detection accuracy.}
\item {We proposed the extended receptive domain (ERD) strategy to capture richer contextual information and reduce false positives caused by nodule occlusion. We also proposed the position channel attention mechanism (PCAM) to optimize feature learning and reduce multiscale detection errors, and designed the tiny object detection block (TODB) to enhance the detection of tiny nodules.}
\item {We conducted thorough experiments on the public LUNA16 dataset \cite{setio2017validation}, achieving state-of-the-art performance, with an mAP improvement of \textbf{8.8\%} over the previous state-of-the-art method YOLOv8.}

\end{itemize}

\section{Related Works}
Dense prediction techniques are widely used in medical imaging, especially for semantic segmentation and object detection \cite{zhang2024meddet}, due to their ability to make pixel-level predictions that improve precision in identifying and localizing anatomical or pathological regions. Deep learning-based pulmonary nodule detection algorithms can be broadly classified into two categories: two-stage and one-stage detection algorithms.

Two-stage algorithms, like R-CNN, first generate candidate regions and then classify and localize targets. Fast R-CNN accelerates detection by sharing convolutional features, while Faster R-CNN improves both speed and accuracy with Region Proposal Networks (RPNs). Xu et al. {\cite{xu2023improved}} enhanced Faster R-CNN with multi-scale training, Online Hard Example Mining (OHEM), customized anchor sizes, and deformable convolutions, improving small target detection. Tong et al. \cite{tong2020pulmonary} combined Faster R-CNN with ISODATA and 3D-CNN for false positive reduction, achieving strong results on LUNA16. However, the main drawback of two-stage methods is their slow detection speed due to the time-consuming candidate region generation, which can also lead to inaccurate localization, especially with varying target shapes and sizes.

One-stage detection algorithms improve on two-stage methods by detecting objects directly from the input image, eliminating candidate region generation and accelerating the detection process. Popular algorithms like YOLO and SSD achieve both localization and classification in a single pass, making them suitable for real-time applications. In pulmonary nodule detection, the YOLO series is widely used due to its speed and accuracy. For example, Wu et al. {\cite{wu2024yolo}} proposed a YOLOv7-based method with Efficient Omni-Dimensional Convolution (EODConv), achieving a mAP of 94.6\% on LUNA16. Zhanlin Ji et al. {\cite{ji2023elct}} improved performance by decoupling the feature pyramid into high- and low-semantic regions, achieving 92.3\% accuracy on the Lung-PETCT-Dx dataset.

However, one-stage algorithms face challenges from down-sampling steps that reduce the feature map size, making small nodules harder to detect. Additionally, occlusion by lung tissues complicates detection, leading to missed detections, false positives, and high false-positive rates. To address these issues, we propose a novel one-stage detection network that integrates multiple modules for more efficient and accurate detection of small, occluded, and multi-scale nodules in complex pulmonary environments.

\section{Methods}
\subsection{Overview }
\begin{figure}[htbp]
\centering
\includegraphics[width=1\linewidth]{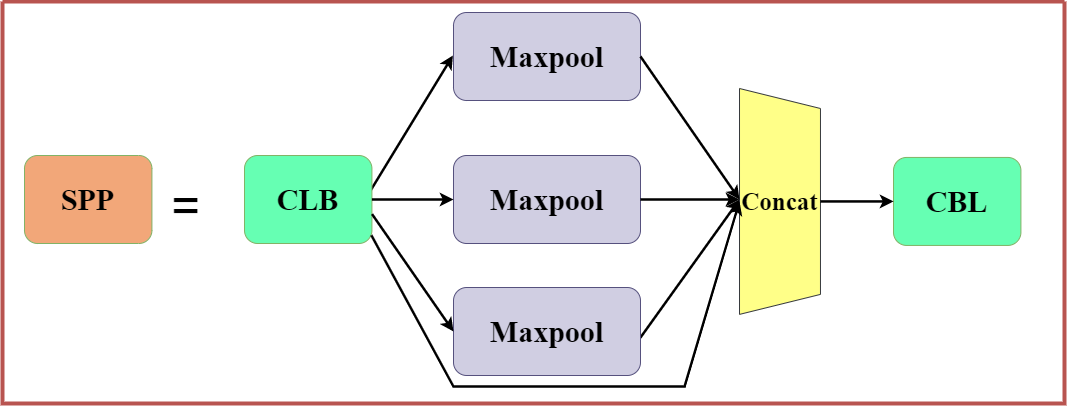} 
\caption{Architecture of the Spatial Pyramid Pooling (SPP) module. The module utilizes a Convolutional Layer Block (CLB) followed by three parallel Maxpool layers with varying sizes to capture multi-scale features. These features are then concatenated and processed by another CLB to enhance the final feature representation, ensuring robust spatial invariance.}
\label{fig:3}
\end{figure}

We propose MSDet, a novel lung nodule detection network (Figure \ref*{fig:2}). MSDet enhances contextual information and reduces false positives from nodule occlusion using the extended receptive domain (ERD) strategy. A position channel attention mechanism (PCAM) optimizes feature learning and reduces multi-scale detection errors. The tiny object detection block (TODB) improves small nodule detection by reducing background interference. Given a lung CT image, MSDet performs convolution and feature fusion, followed by multi-scale fusion and information transfer. The TODB restores the feature map size via upsampling, which is then fused with different resolution feature maps for refined predictions. The ERD module expands the receptive field, captures more contextual details, and integrates information from a larger area. As shown in {Figure \ref*{fig:3}}, the SPP module further strengthens this by aggregating features at multiple scales, improving robustness. In the PCAM module, position attention captures long-range context, while channel attention models interdependencies between channels, adjusting weights based on importance. Finally, the model generates multi-scale feature maps for the final predictions.

\subsection{Tiny Object Detection Block (TODB)}
\begin{figure}[htbp]
\centering
\includegraphics[width=1\linewidth]{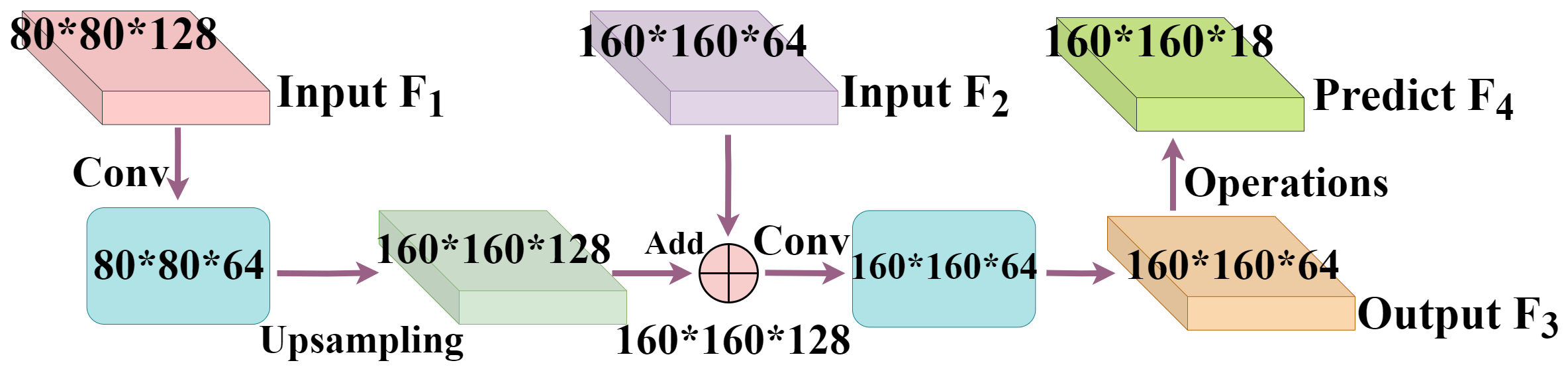} 
\caption{TODB Structure. This module integrates multi-resolution features through upsampling and feature map fusion, allowing the network to capture small pulmonary nodules more accurately. The structure enhances detection robustness by combining features from different resolutions.}
\label{fig:4}
\end{figure}

Vanilla YOLOv5, designed for natural images, struggles with detecting pulmonary nodules in CT scans due to excessive downsampling, which reduces feature map size and makes small nodules nearly undetectable. For instance, 32x downsampling reduces a 20×20 pixel nodule to just 1×1 pixel. To address this, we introduced the TODB module, which captures relationships between feature maps and minimizes interference from surrounding tissues. By optimizing feature map connections and enhancing information extraction, TODB improves the detection accuracy of small nodules, as shown in Figure \ref*{fig:4}.

The network input size is 640×640×3, and after a series of downsampling operations, an 80×80×128 feature map \( F_1 \) is generated. To enrich it with contextual information, a 1×1 convolution is applied, altering the channel number for cross-channel interaction, enhancing nodule detection. This operation is expressed as:

\begin{equation}
F_1' = \sigma(W_1 * F_1)
\end{equation}

where \( \sigma \) is the activation function and \( * \) denotes convolution. Next, \( F_1 \) is upsampled to \( 160 \times 160 \times 64 \) and added to another feature map \( F_2 \), obtained via a 4x downsampling of the input. This operation combines multi-resolution features to improve detection accuracy, expressed as:

\begin{equation}
F_3 = \text{Upsample}(F_1') + F_2
\end{equation}

After further convolutional processing, the final output \( F_4 \) is generated, which is a \( 160 \times 160 \times 18 \) feature map used for nodule prediction:

\begin{equation}
F_4 = \sigma(W_2 * F_3)
\end{equation}

This process divides the input into \( 160 \times 160 \) regions for precise nodule prediction, efficiently integrating multi-resolution and cross-channel information to enhance detection accuracy. By concentrating on smaller regions and leveraging multi-scale features, the model's robustness is greatly enhanced.

\subsection{Extended Receptive Domain (ERD)}
\begin{figure}[htbp]
\centering
\includegraphics[width=0.5\linewidth]{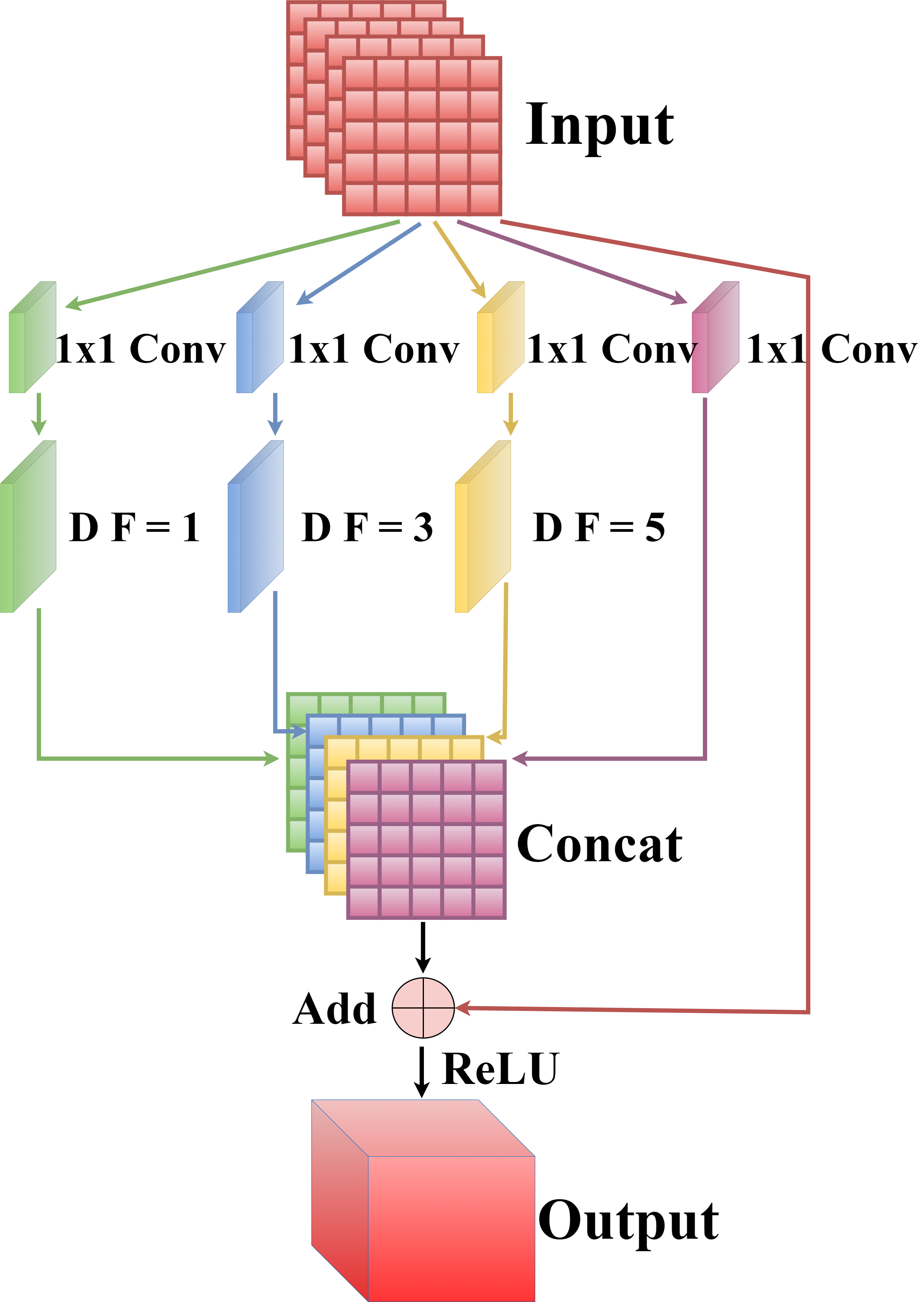} 
\caption{ERD Structure. The ERD consists of multiple parallel convolution branches with varying dilation rates, designed to capture features at different spatial scales. This structure enhances the detection of lung nodules by expanding the receptive field while maintaining computational efficiency.}
\label{fig:6}
\end{figure}

Decoupling network bottleneck structures can improve lung nodule detection but may reduce the receptive field, limiting global context. To address this, we propose an extended receptive domain (ERD) strategy, which expands the receptive field while maintaining efficiency, enhancing detection across various nodule sizes. As shown in Figure \ref{fig:6}, ERD uses a multi-branch structure with three parallel 3×3 dilated convolution branches (dilation rates \( R = (1, 3, 5) \)), a 1×1 convolution branch, and an identity branch. This approach provides varying receptive fields, captures fine details, and preserves positional information.

\begin{figure}[htbp]
\centering
\includegraphics[width=1\linewidth]{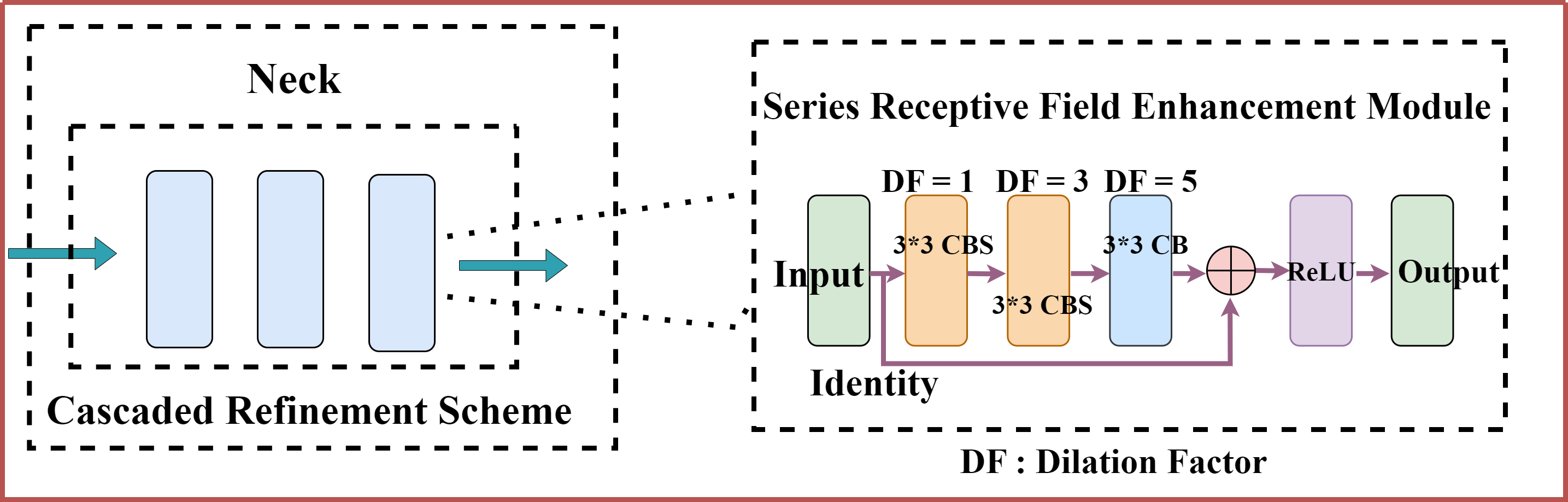} 
\caption{Illustration of the ERD architecture integrating multiple dilated convolutions for lung nodule detection. The diagram on the left shows the Neck with a Cascaded Refinement Scheme, and the right side details the Series Receptive Field Enhancement Module, employing dilated convolutions with varying dilation factors to capture multiscale features effectively.}
\label{fig:5}
\end{figure}

Dilated convolutions introduce fixed gaps between kernel values, allowing expansion of the receptive field without increasing the number of parameters. The output of a dilated convolution with input \( x(m,n) \) and kernel weights \( \omega(i,j) \) can be expressed as:

\begin{equation}
y(m,n) = \sum_{i=1}^{M} \sum_{j=1}^{N} x(m + r \cdot i + n + r \cdot j) \omega(i,j)
\end{equation}

where \( r \) is the dilation rate, and \( M \), \( N \) are kernel sizes (typically 3×3). By varying \( r \), the receptive field expands: \( r=1 \) results in a 3×3 receptive field, \( r=2 \) gives 5×5, and \( r=3 \) gives 7×7, all while keeping the computation similar to a standard convolution.

As illustrated in {Figure \ref*{fig:5}}, the ERD strategy employs a sophisticated multi-branch structure to effectively capture features at multiple scales. This figure visually demonstrates the arrangement and interaction of various convolution branches within the ERD, highlighting the specific roles of dilated and 1×1 convolutions in enhancing the model's sensitivity to spatial details and contextual variations across different nodule sizes.

The equivalent receptive field (RF) for a \( k \times k \) dilated convolution and the output feature map resolution (H) are calculated as:

\begin{equation}
\text{RF} = (r-1)(k-1) + k
\end{equation}
\begin{equation}
H = \frac{(h + 2p - \text{RF})}{s} + 1
\end{equation}

where \( p \) is the padding, \( h \) is the input feature map resolution, and \( s \) is the stride. This method enables efficient scaling of the receptive field without increasing computational cost or altering the spatial resolution of the feature maps.

\subsection{Position Channel Attention Mechanism (PCAM)}
\begin{figure}[htbp]
\centering
\includegraphics[width=1\linewidth]{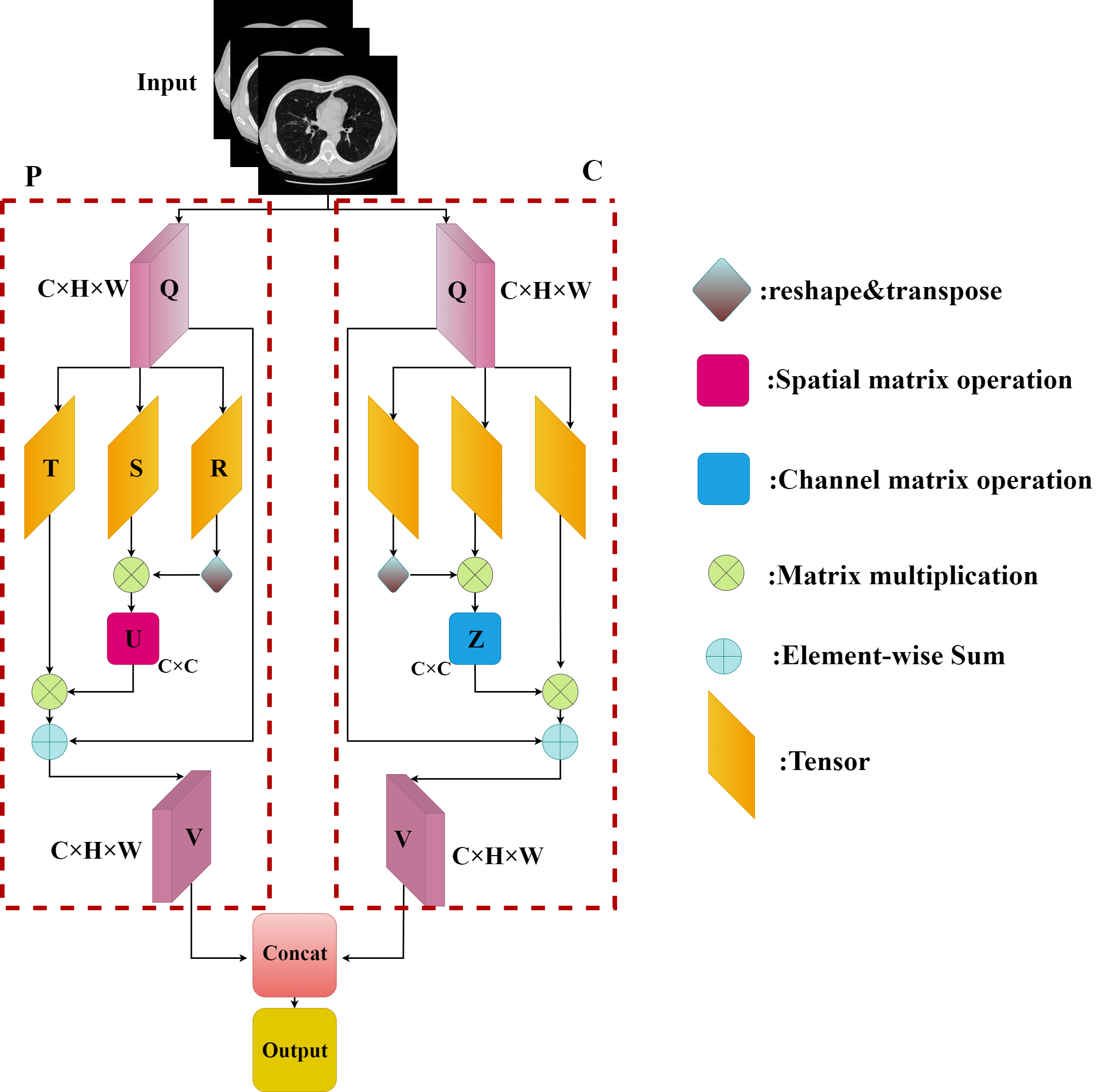} 
\caption{PCAM Structure. The diagram illustrates the PCAM employing a dual-module approach to capture and integrate complex feature interactions through spatial and channel attentions. This innovative mechanism leverages higher-order statistics to enhance activation feature characterization, optimizing the performance for intricate image analysis challenges.}
\label{fig:7}
\end{figure}

Attention mechanisms in CNNs highlight key features while suppressing irrelevant ones, improving performance. Existing methods rely on first-order statistics, limiting feature interaction capture. To address this, we propose PCAM (Figure \ref{fig:7}), which incorporates higher-order statistics for better feature representation. Applied at the MSDet bottleneck, PCAM enhances focus on key regions in pulmonary nodule images. It consists of a positional attention module, which captures long-range spatial dependencies by encoding contextual information. Given a feature map $Q \in \mathbb{R}^{C \times H \times W}$, we generate two feature maps, $R$ and $S$, and compute a spatial attention map $U \in \mathbb{R}^{N \times N}$ as follows:
\begin{equation}
u_{ji} = \frac{e^{(R_i \cdot S_j)}}{\sum_{i=1}^{N} e^{(R_i \cdot S_j)}}
\end{equation}

Here, $u_{ji}$ represents the correlation between positions $i$ and $j$. The resulting attention-weighted feature map is then added to the original features, scaled by a learned parameter $\beta$:

\begin{equation}
V_j = \beta \sum_{i=1}^{N} (u_{ji} T_i) + Q_j
\end{equation}

Similarly, the channel attention module models dependencies between feature channels. Given $Q$, the channel attention map $Z \in \mathbb{R}^{C \times C}$ is computed as:

\begin{equation}
z_{ji} = \frac{e^{(Q_i \cdot Q_j)}}{\sum_{i=1}^{C} e^{(Q_i \cdot Q_j)}}
\end{equation}

The final feature map is computed as:

\begin{equation}
V_j = \gamma \sum_{i=1}^{C} (z_{ji} Q_i) + Q_j 
\end{equation}

Both $\beta$ and $\gamma$ are initialized to 0 and learn appropriate weights during training. By combining positional and channel attention, PCAM effectively enhances feature discriminability.

\section{Experiments}
\subsection{Dataset and Evaluation Matrices}
The experimental data were obtained from the publicly available LUNA16 lung CT dataset, with detailed dataset configurations provided in \textbf{Supplementary Section 1}. The model performance was evaluated using standard metrics such as precision, recall, F1-score, and mAP.

\subsection{Implementation Details}
\begin{figure}[htbp]
\centering
\includegraphics[width=1\linewidth]{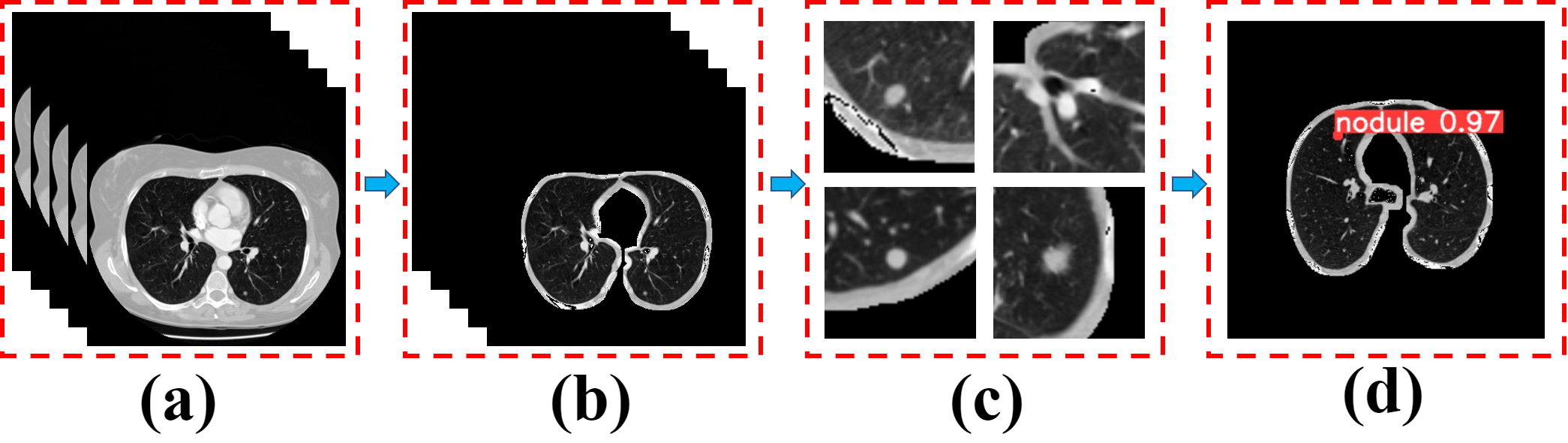} 
\caption{Illustration of the pulmonary nodule detection process using the MSDet model. The sequence shows (a) the initial input CT image, (b) processed two-dimensional slices after image preprocessing, (c) zoomed-in views highlighting candidate nodules after candidate detection, and (d) the final detection results with a confidence score.}
\label{fig:8}
\end{figure}

The MSDet model was implemented using PyTorch 1.13.1 and trained on an NVIDIA GPU with a batch size of 8 for 200 epochs. Detailed experimental parameters and configurations are provided in \textbf{Supplementary Section 2}. The detection workflow is shown in Figure \ref{fig:8}.

\subsection{Comparative Studies}
We performed comparative analyses on the LUNA16 dataset, as shown in Table \ref{table:3}, against state-of-the-art methods. MSDet significantly outperforms leading models, achieving a mAP of 97.3\%, surpassing Faster R-CNN by 6.6 percentage points and YOLOv5 by 11.5\%. Additionally, MSDet outperformed EfficientDet-d and CenterNet by 9.9\% and 12\%, respectively. These results highlight MSDet's superior detection accuracy and efficiency, validating its effectiveness for lung nodule detection.

\subsection{Ablation Studies}
The ablation study in \textbf{Supplementary Table~(1)} shows the significant improvements achieved by TODB, ERD, and PCAM in the model’s performance on the LUNA16 dataset. The combination of these modules led to a substantial boost in precision, recall, and mAP\textsubscript{0.5}, with final metrics reaching 97.2\%, 96.1\%, and 97.3\%, respectively. \textbf{Supplementary Table~(2)} compares various attention mechanisms, revealing that PCAM outperforms others in precision, recall, and F1 score, while enhancing multi-scale detection and reducing false positives. For detailed results, please refer to \textbf{Supplementary, Section 3}.

\begin{table}[htbp]
\centering
\caption{Comparison of MSDet with state-of-the-art lung nodule detection networks in terms of mAP. The table highlights the superior performance of MSDet compared to widely-used models such as YOLO, Faster R-CNN, and LungSeek.}
\fontsize{8}{11}\selectfont
\setlength{\tabcolsep}{25pt}
\label{table:3}
\begin{tabular}{lcccc}
\hline
Model & mAP \\
\hline
YOLOv5{\cite{ega2023study}} & 85.8\% \\
YOLOv6{\cite{goel2024improving}} & 82.79\% \\
YOLOv7{\cite{mammeri2024early}} & 81.28\% \\
YOLOv8{\cite{csaman2023yolov8}} & 88.5\% \\
Faster R-CNN{\cite{xu2023improved}} & 90.7\% \\
Two-Stage CNN{\cite{jain2023pulmonary}} & 84.4\% \\
LungSeek{\cite{zhang2023lungseek}} & 91.75\% \\
DLDS-CT{\cite{lu2024dual}} & 81.2\% \\
EfficientDet-d{\cite{tan2020efficientdet}} & 87.4\% \\
CenterNet{\cite{zhou2019objects}} & 85.3\% \\
STBi-YOLO{\cite{liu2022stbi}} & 95.9\% \\
YOLOv5-CASP{\cite{ji2023lung}} & 72.0\% \\
YOLO-MSRF{\cite{wu2024yolo}} & 94.6\% \\
\midrule
\textbf{MSDet (Ours)} & \textbf{97.3\%} \\
\hline
\end{tabular}
\end{table}

\subsection{Discussion}
\subsubsection{Clinical Impact}
MSDet improves accuracy and reduces false positives, enhancing early malignancy detection. Its automated, lightweight design eases radiologists' workload and supports real-time deployment, making it ideal for large-scale screenings and alleviating healthcare system burdens.

\subsubsection{Contributions to Early Diagnosis and Broader Societal Impact}
MSDet enables earlier lung cancer detection, improving survival rates and reducing invasive treatments. By minimizing false positives, it cuts follow-up costs and stress, while enhancing diagnostic efficiency, especially in underserved areas, ultimately reducing healthcare burdens.

\subsection{Visualization}
\begin{figure}
  \begin{minipage}[h]{\linewidth}
    \centering
    \includegraphics[width=1\textwidth]{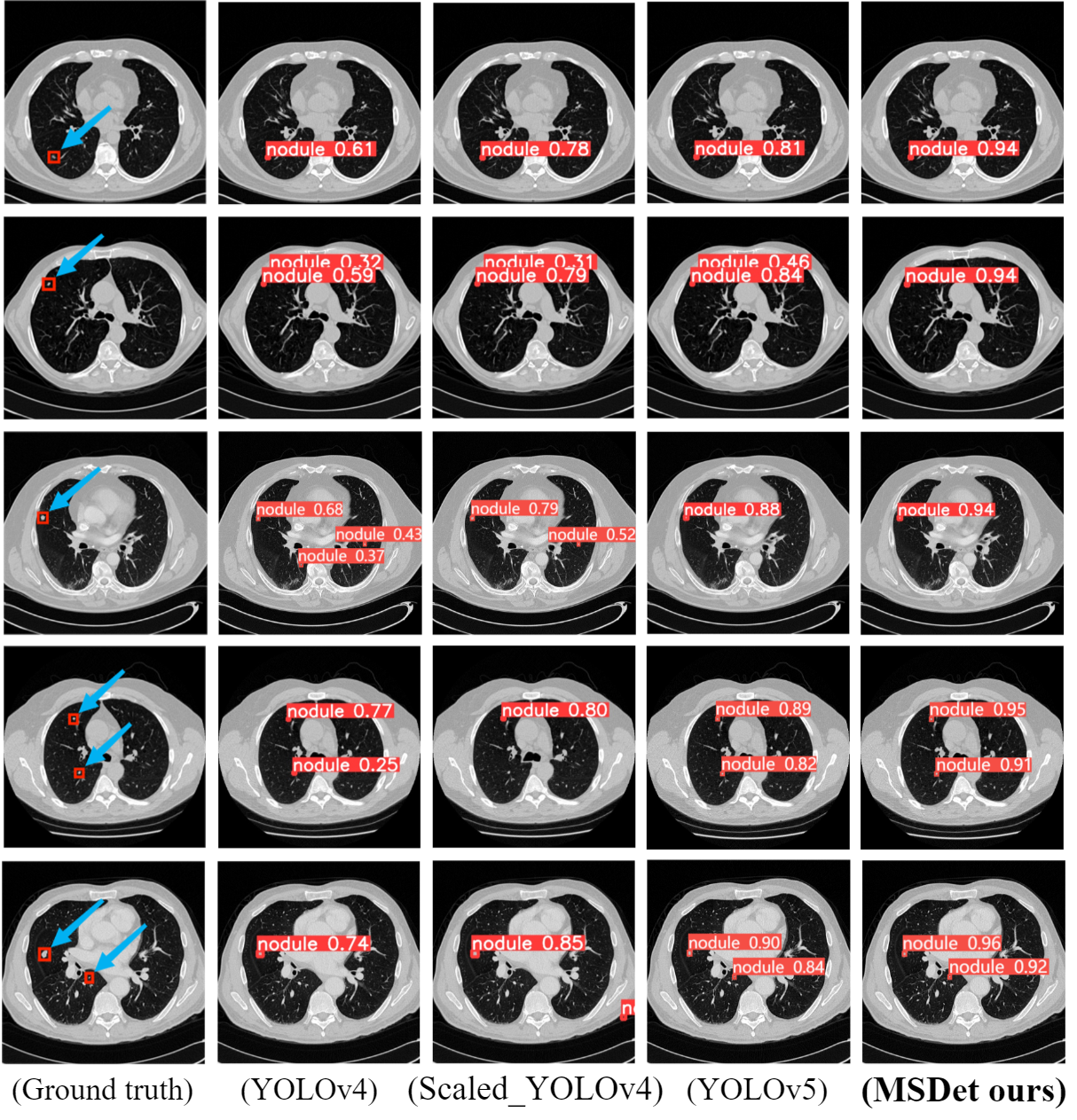}
    \caption{Visual comparison of lung nodule detection results on the LUNA16 dataset, highlighting MSDet's superior performance compared with other models. The figure illustrates MSDet's effectiveness in reducing false positives and achieving higher accuracy, showcasing its robustness for clinical applications.}
    \label{fig:9}
  \end{minipage}
\end{figure}

\begin{figure}
  \begin{minipage}[h]{\linewidth}
    \centering
    \includegraphics[width=1\textwidth]{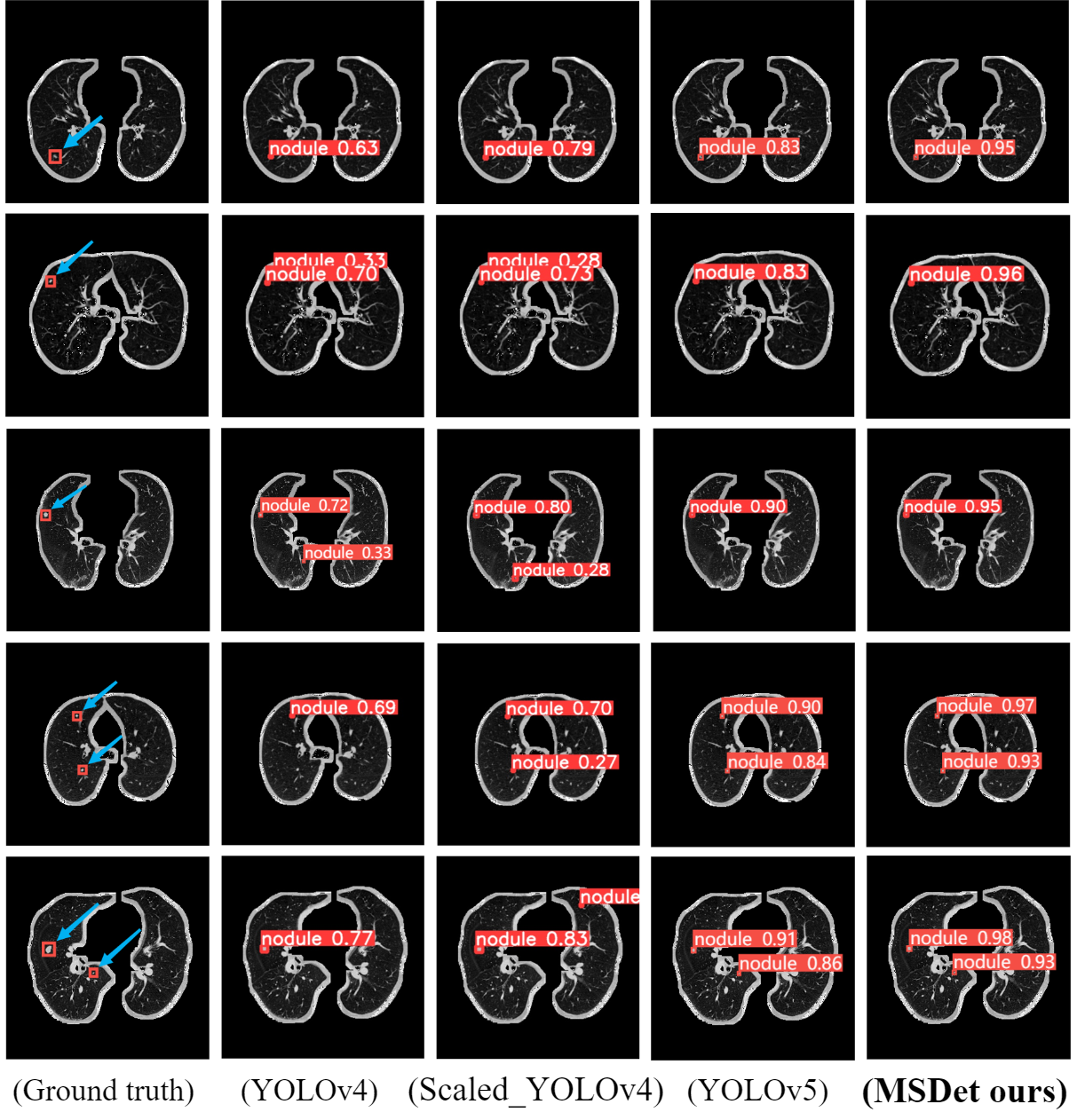}
    \caption{Visual comparison of lung nodule detection on the LUNA16 dataset, showing the impact of lung parenchyma segmentation. MSDet achieves up to 98\% accuracy, reducing false positives and improving detection precision in complex scenarios.}
    \label{fig:10}
  \end{minipage}
\end{figure}

We validate the effectiveness of MSDet for lung nodule detection through visual comparison. As shown in Figure \ref{fig:9}, MSDet outperforms YOLOv4, Scaled\_YOLOv4, and the Basic Model in both single and multi-nodule tasks, achieving 94\% accuracy in complex backgrounds. It identifies all nodules with no false positives or negatives, outperforming other models by 6-8\%. Lung parenchyma segmentation further boosts accuracy, as shown in Figure \ref{fig:10}, enabling MSDet to achieve up to 98\% accuracy and significantly reduce false detections. These results demonstrate the robustness of MSDet and the importance of segmentation for improving performance in complex imaging conditions, making it well-suited for clinical applications.

\section{Conclusion}
This paper presents MSDet, a novel method for lung nodule detection in CT images, designed to address multi-scale detection challenges. By designed advanced strategies such as TODB, ERD, and PCAM, MSDet enhances detection accuracy, especially for small and occluded nodules. Our model achieves superior performance compared to existing methods, demonstrating a significant improvement in detecting nodules in complex backgrounds. Empirical results on the LUNA16 dataset validate that MSDet outperforms current benchmarks in both precision and recall. This work marks a step forward in medical image analysis, offering valuable insights for future advancements in clinical detection systems.


\clearpage

\input{supplementary}

\clearpage

\input{icme2025_template_anonymized.bbl}

\end{document}

%% file: supplementary.tex
\newcommand{\supplementarytitle}{
    \twocolumn[
        \begin{center}
            \Large \textbf{Supplementary Material}
            \vspace{1em} 
        \end{center}
    ]
}

\supplementarytitle

\section{Dataset and Evaluation Matrices}

The experimental data utilized in this study were obtained from the publicly available LUNA16 lung CT imaging dataset. The LUNA16 dataset is derived from the LIDC-IDRI dataset, with slices thicker than 3 mm and nodules smaller than 3 mm removed, encompassing a total of 888 cases. To facilitate analysis, we converted the three-dimensional data into two-dimensional slices.

In terms of evaluation metrics, the performance of the pulmonary nodule detection model is assessed using several standard measures. Precision quantifies the accuracy of positively predicted instances, aiming to minimize false positives. Recall evaluates the model's ability to identify all relevant cases, which is crucial for thorough detection. The F1-score merges precision and recall in a single metric, reflecting the balance between detection accuracy and completeness. Average Precision (AP), calculated over the Precision-Recall curve, provides detailed insights into the model's precision across various recall levels. Mean Average Precision (mAP) averages these AP values across different object classes and sizes, offering a global view of the model’s overall accuracy.

\section{Specific preprocessing steps}
The proposed MSDet model was implemented using the PyTorch 1.13.1 framework and trained on an NVIDIA GPU. The batch size was set to 8, and training spanned 200 epochs. We employed the SGD optimizer with a momentum of 0.937 and an initial learning rate of 0.01. Data augmentation techniques, including random rotations, flips, brightness, contrast, and color adjustments, along with salt-and-pepper noise, were applied to enhance model robustness.

Prior to training, the data underwent a comprehensive preprocessing phase. The CT image grayscale values were first converted to Hounsfield Units (HU), which reflect the radiodensity of human tissues, with lung tissue typically around -500 HU. During preprocessing, we retained regions with HU values between -1000 and 400 while truncating values outside this range. The raw data were then clipped to a range of [-1200, 600], setting values below -1200 and above 600 to -1200 and 600 respectively, effectively filtering out elements like water and air. Lung parenchyma segmentation was performed using erosion to remove granular regions, followed by dilation to encompass blood vessels and eliminate black noise from non-transparent rays within the lung regions. The central part of the CT image was selected for lung mask extraction, focusing on the largest connected component, which was further dilated to fully capture the lung region. If the edges remained uneven, additional erosion was applied to refine them. Finally, the preprocessed images were normalized to a range of 0 to 255, completing the data preprocessing pipeline.

After preprocessing, the performance of the models was evaluated using standard metrics such as mAP\textsubscript{0.5} and mAP at 0.5 to 0.95. MSDet consistently outperformed models like YOLOv3, YOLOv7, Scaled\_YOLOv4, and TPH-YOLOv5, achieving a 60\% mAP at 0.5 to 0.95 after 125 epochs, demonstrating higher precision and recall with fewer training iterations.

The lung nodule detection workflow is depicted in Figure 8.

\section{Supplementary Materials – Ablation Study Results}

The detailed results of the ablation study are provided in this section. As shown in Supplementary Table \ref{table:1}, we present a comprehensive analysis of the model’s performance across different configurations. The study evaluates the impact of various modules, including TODB, ERD, and PCAM, on the model’s precision, recall, and mAP\textsubscript{0.5}. These results highlight the contribution of each module and demonstrate the significant improvements achieved when all components are integrated. Specifically, the combination of TODB, ERD, and PCAM yields the highest performance, underscoring their importance in enhancing the model’s detection capabilities.

\begin{table}[htbp]
\centering
\caption{Ablation study of MSDet on the LUNA16 dataset, evaluating the impact of different modules (TODB, ERD, and PCAM) on detection performance.}
\fontsize{8.4}{8}\selectfont 
\setlength{\tabcolsep}{2.6pt} 
\label{tab:ap_excellent_algorithms}
\begin{tabular}{lccc|cccc}
\toprule
YOLOv5 & TODB & ERD & PCAM & Precision (\%) & Recall (\%) & mAP\textsubscript{0.5} (\%) \\
\midrule
\checkmark &  &  &  & 83.7 ± 0.09 & 81.3 ± 0.07 & 84.7 ± 0.03 \\
\checkmark & \checkmark &  &  & 88.4 ± 0.04 & 87.2 ± 0.05 & 89.1 ± 0.08 \\
\checkmark &  & \checkmark &  & 88.5 ± 0.02 & 87.6 ± 0.06 & 88.9 ± 0.04 \\
\checkmark &  &  & \checkmark & 87.1 ± 0.09 & 88.7 ± 0.05 & 88.3 ± 0.07 \\
\checkmark & \checkmark & \checkmark &  & 91.6 ± 0.08 & 90.1 ± 0.06 & 92.8 ± 0.02 \\
\checkmark & \checkmark &  & \checkmark & 91.4 ± 0.04 & 91.2 ± 0.03 & 92.5 ± 0.08 \\
\checkmark &  & \checkmark & \checkmark & 91.2 ± 0.09 & 91.4 ± 0.05 & 91.9 ± 0.01 \\
\midrule
\checkmark & \checkmark & \checkmark & \checkmark & \textbf{97.2 }± 0.06 & \textbf{96.1 }± 0.05 & \textbf{97.3 }± 0.03 \\
\bottomrule
\end{tabular}
  \label{table:1}
\end{table}

\begin{table*}[t]
\centering
\caption{Detection performance comparison using different attention modules on the LUNA16 dataset. This table presents precision, recall, F1 score, and mAP results across various IoU thresholds and nodule sizes, demonstrating the superior performance of the PCAM (Ours) module compared to other state-of-the-art attention mechanisms.}
\fontsize{7.7}{8.3}\selectfont 
\setlength{\tabcolsep}{6pt} 
\label{tab:ap_excellent_algorithms}
\begin{tabular}{l|ccc|ccc|ccc}
\toprule
Methods & Precision(\%) & Recall(\%) & F1\textsubscript{Score}(\%) & mAP\textsubscript{0.5} (\%) & mAP\textsubscript{0.75} (\%) & mAP\textsubscript{0.95} (\%) & mAP\textsubscript{S} (\%) & mAP\textsubscript{M} (\%) & mAP\textsubscript{L} (\%) \\
\midrule
SE{\cite{hu2018squeeze}}   & \num{89.5} $\pm$ \num{0.02} & 90.1 $\pm$ 0.04 & 89.8 $\pm$ 0.06 & 89.7 $\pm$ 0.03 & 79.2 $\pm$ 0.08 & 60.7 $\pm$ 0.09 & 67.4 $\pm$ 0.1 & 74.1 $\pm$ 0.02 & 80.1 $\pm$ 0.08 \\
CBAM{\cite{woo2018cbam}} & \num{91.4} $\pm$ \num{0.1} & 91.9 $\pm$ 0.03 & 91.6 $\pm$ 0.08 & 91.7 $\pm$ 0.03 & 81.1 $\pm$ 0.1 & 63.2 $\pm$ 0.04 & 68.1 $\pm$ 0.07 & 75.2 $\pm$ 0.05 & 81.4 $\pm$ 0.02 \\
SA{\cite{zhang2021sa}}   & \num{89.8} $\pm$ \num{0.06} & 90.6 $\pm$ 0.07 & 90.2 $\pm$ 0.01 & 90.2 $\pm$ 0.09 & 79.8 $\pm$ 0.08 & 61.1 $\pm$ 0.03 & 67.1 $\pm$ 0.03 & 74.0 $\pm$ 0.06 & 80.5 $\pm$ 0.01 \\
NAM{\cite{liu2021nam}}  & \num{88.6} $\pm$ \num{0.1} & 89.1 $\pm$ 0.06 & 88.8 $\pm$ 0.03 & 88.4 $\pm$ 0.02 & 79.1 $\pm$ 0.08 & 60.4 $\pm$ 0.08 & 67.3 $\pm$ 0.09 & 74.7 $\pm$ 0.04 & 80.3 $\pm$ 0.07 \\
CA{\cite{hou2021coordinate}}   & \num{90.3} $\pm$ \num{0.04} & 89.7 $\pm$ 0.07 & 89.9 $\pm$ 0.08 & 90.1 $\pm$ 0.09 & 79.4 $\pm$ 0.03 & 61.6 $\pm$ 0.04 & 67.6 $\pm$ 0.02 & 75.1 $\pm$ 0.05 & 81.0 $\pm$ 0.09 \\
ECA{\cite{wang2020eca}}  & \num{89.5} $\pm$ \num{0.09} & 90.2 $\pm$ 0.06 & 89.8 $\pm$ 0.06 & 89.7 $\pm$ 0.03 & 79.6 $\pm$ 0.07 & 60.8 $\pm$ 0.02 & 67.9 $\pm$ 0.08 & 74.1 $\pm$ 0.1 & 80.2 $\pm$ 0.1 \\
NLNN{\cite{wang2018non}} & 90.4 $\pm$ 0.1 & 90.6 $\pm$ 0.03 & 89.6 $\pm$ 0.05 & 89.2 $\pm$ 0.02 & 79.4 $\pm$ 0.07 & 61.2 $\pm$ 0.09 & 67.2 $\pm$ 0.05 & 74.9 $\pm$ 0.04 & 81.1 $\pm$ 0.08 \\
GAT{\cite{velivckovic2017graph}}  & 89.2 $\pm$ 0.03 & 89.7 $\pm$ 0.01 & 89.4 $\pm$ 0.05 & 89.8 $\pm$ 0.09 & 80.5 $\pm$ 0.1 & 60.7 $\pm$ 0.07 & 67.0 $\pm$ 0.04 & 74.6 $\pm$ 0.02 & 80.7 $\pm$ 0.08 \\
BAM{\cite{park2018bam}}  & 89.8 $\pm$ 0.07 & 90.2 $\pm$ 0.08 & 90.2 $\pm$ 0.07 & 90.5 $\pm$ 0.04 & 80.1 $\pm$ 0.02 & 61.4 $\pm$ 0.06 & 67.9 $\pm$ 0.1 & 74.8 $\pm$ 0.05 & 81.0 $\pm$ 0.01 \\
MAB{\cite{vaswani2017attention}}  & 90.9 $\pm$ 0.02 & 91.2 $\pm$ 0.05 & 91.1 $\pm$ 0.03 & 91.3 $\pm$ 0.06 & 79.9 $\pm$ 0.09 & 61.9 $\pm$ 0.08 & 68.1 $\pm$ 0.02 & 75.1 $\pm$ 0.03 & 80.2 $\pm$ 0.1 \\
CCA{\cite{huang2019ccnet}}  & \num{91.4} $\pm$ \num{0.1} & 91.8 $\pm$ 0.07 & 91.6 $\pm$ 0.02 & 91.5 $\pm$ 0.09 & 80.2 $\pm$ 0.01 & 62.3 $\pm$ 0.03 & 68.0 $\pm$ 0.05 & 74.4 $\pm$ 0.04 & 80.4 $\pm$ 0.06 \\
SOCA{\cite{dai2019second}} & 90.7 $\pm$ 0.07 & 90.3 $\pm$ 0.06 & 90.5 $\pm$ 0.02 & 90.3 $\pm$ 0.1 & 80.6 $\pm$ 0.03 & 61.7 $\pm$ 0.1 & 68.2 $\pm$ 0.04 & 75.0 $\pm$ 0.08 & 81.5 $\pm$ 0.04 \\
\midrule
\textbf{PCAM (Ours)} & \textbf{93.1} $\pm$ \num{0.05} & \textbf{92.5} $\pm$ \num{0.04} & \textbf{92.8} $\pm$ \num{0.07} & \textbf{93.7} $\pm$ \num{0.05} & \textbf{83.7} $\pm$ \num{0.1} & \textbf{65.8} $\pm$ \num{0.03} & \textbf{70.6} $\pm$ \num{0.03} & \textbf{77.1} $\pm$ \num{0.06} & \textbf{83.7} $\pm$ \num{0.07} \\
\bottomrule
\end{tabular}
  \label{table:2}
\end{table*}

Furthermore, Supplementary Table \ref{table:2} presents a comparison of different attention mechanisms on the LUNA16 dataset, including SE, CBAM, SA, NAM, CA, ECA, NLNN, GAT, BAM, MAB, CCA, SOCA, and our proposed PCAM module. The table reports precision, recall, F1 score, and mAP results across various IoU thresholds and nodule sizes. The results clearly demonstrate that PCAM outperforms other attention mechanisms in all metrics, achieving the highest precision (93.1\%), recall (92.5\%), F1 score (92.8\%), and mAP\textsubscript{0.5} (93.7\%) values. Additionally, PCAM consistently shows superior performance across different nodule sizes (mAP\textsubscript{S}, mAP\textsubscript{M}, and mAP\textsubscript{L}), reinforcing its effectiveness in improving multi-scale detection and reducing false positives.

\section{Detailed Analysis of Training Progress}
\begin{figure}[htbp]
\centering
\includegraphics[width=\linewidth]{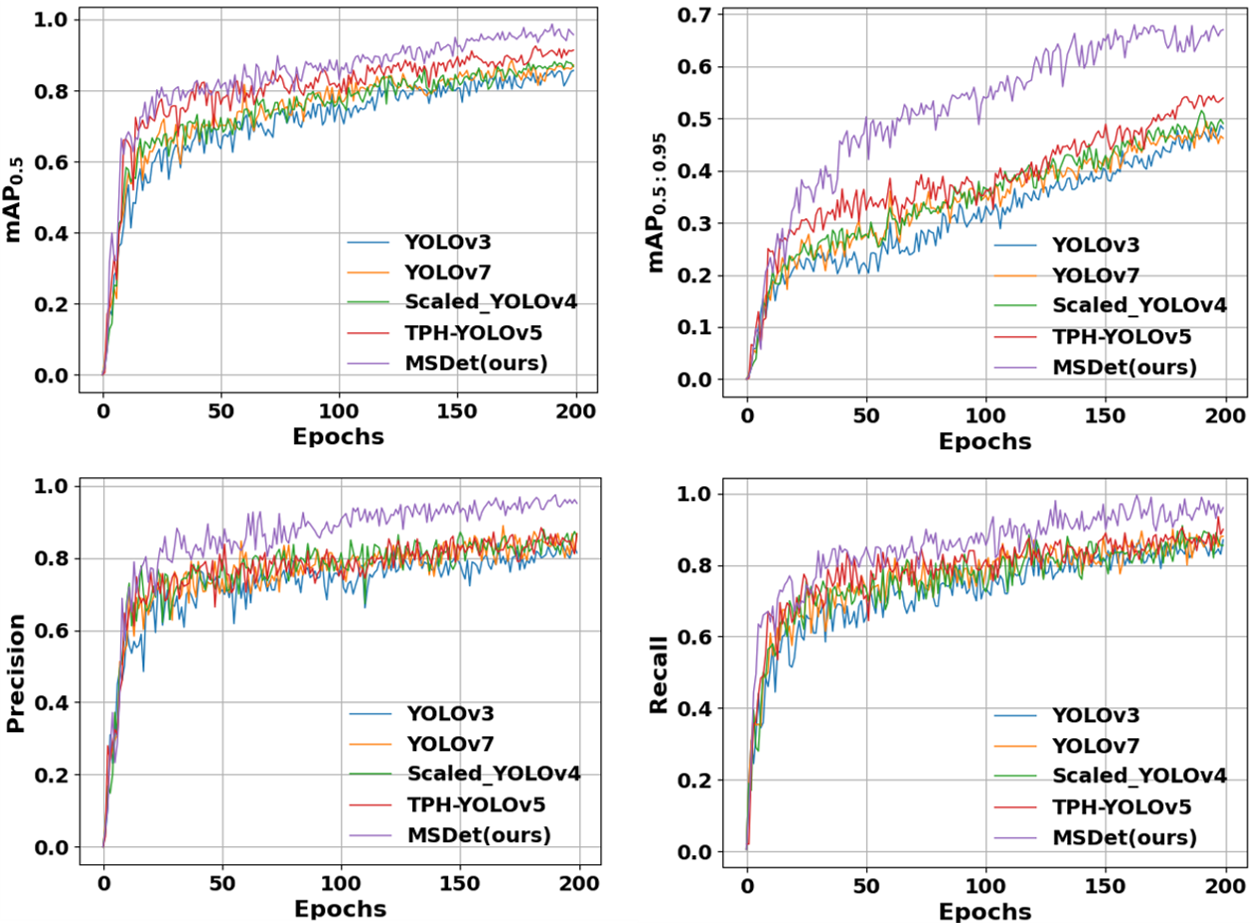} 
\caption{Training performance comparison of MSDet with YOLO variants across 200 epochs. The graph demonstrates MSDet's superior performance with a clear lead in mAP scores from IoU 0.5 to 0.95, underscoring its rapid convergence and higher accuracy in detecting more complex features effectively.}
\label{fig:12}
\end{figure}

The training curves shown in Figure \ref{fig:12} compare the performance of the MSDet model against YOLOv3, YOLOv7, Scaled\_YOLOv4, and TPH-YOLOv5 across 200 epochs using metrics such as mAP\textsubscript{0.5} and mAP from 0.5 to 0.95. MSDet demonstrates rapid convergence and superior performance, achieving a 60\% mAP at 0.5 to 0.95 by epoch 125, highlighting its efficient learning and adaptation capabilities.

The graph presents key metrics including mAP at IoU 0.5, mAP from 0.5 to 0.95, precision, and recall. These metrics illustrate MSDet’s consistently higher precision and recall, maintaining smooth and stable progress. The model's early high performance, especially in mAP at IoU 0.5, underscores its effective adaptation to complex image data with fewer epochs compared to other models.This enhanced performance is due to MSDet’s sophisticated feature extraction and optimized learning strategies. 

%% file: icme2025_template_anonymized.bbl
\begin{thebibliography}{10}
\expandafter\ifx\csname url\endcsname\relax
  \def\url#1{\texttt{#1}}\fi
\expandafter\ifx\csname urlprefix\endcsname\relax\def\urlprefix{URL }\fi
\expandafter\ifx\csname href\endcsname\relax
  \def\href#1#2{#2} \def\path#1{#1}\fi

\bibitem{siegel2024cancer}
R.~L. Siegel, A.~N. Giaquinto, A.~Jemal, Cancer statistics, 2024, CA: a cancer journal for clinicians 74~(1) (2024) 12--49.

\bibitem{girshick2014rich}
R.~Girshick, J.~Donahue, T.~Darrell, J.~Malik, Rich feature hierarchies for accurate object detection and semantic segmentation, in: Proceedings of the IEEE conference on computer vision and pattern recognition, 2014, pp. 580--587.

\bibitem{redmon2016you}
J.~Redmon, S.~Divvala, R.~Girshick, A.~Farhadi, You only look once: Unified, real-time object detection, in: Proceedings of the IEEE conference on computer vision and pattern recognition, 2016, pp. 779--788.

\bibitem{vaswani2017attention}
A.~Vaswani, N.~Shazeer, N.~Parmar, J.~Uszkoreit, L.~Jones, A.~N. Gomez, {\L}.~Kaiser, I.~Polosukhin, Attention is all you need, Advances in neural information processing systems 30 (2017).

\bibitem{setio2017validation}
A.~A.~A. Setio, A.~Traverso, T.~De~Bel, M.~S. Berens, C.~Van Den~Bogaard, P.~Cerello, H.~Chen, Q.~Dou, M.~E. Fantacci, B.~Geurts, et~al., Validation, comparison, and combination of algorithms for automatic detection of pulmonary nodules in computed tomography images: the luna16 challenge, Medical image analysis 42 (2017) 1--13.

\bibitem{zhang2024meddet}
Z.~Zhang, N.~Yi, S.~Tan, Y.~Cai, Y.~Yang, L.~Xu, Q.~Li, Z.~Yi, D.~Ergu, Y.~Zhao, Meddet: Generative adversarial distillation for efficient cervical disc herniation detection, arXiv preprint arXiv:2409.00204 (2024).

\bibitem{xu2023improved}
J.~Xu, H.~Ren, S.~Cai, X.~Zhang, An improved faster r-cnn algorithm for assisted detection of lung nodules, Computers In Biology And Medicine 153 (2023) 106470.

\bibitem{tong2020pulmonary}
C.~Tong, B.~Liang, M.~Zhang, R.~Chen, A.~K. Sangaiah, Z.~Zheng, T.~Wan, C.~Yue, X.~Yang, Pulmonary nodule detection based on isodata-improved faster rcnn and 3d-cnn with focal loss, ACM Transactions on Multimedia Computing, Communications, and Applications (TOMM) 16~(1s) (2020) 1--9.

\bibitem{wu2024yolo}
X.~Wu, H.~Zhang, J.~Sun, S.~Wang, Y.~Zhang, Yolo-msrf for lung nodule detection, Biomedical Signal Processing and Control 94 (2024) 106318.

\bibitem{ji2023elct}
Z.~Ji, J.~Zhao, J.~Liu, X.~Zeng, H.~Zhang, X.~Zhang, I.~Ganchev, Elct-yolo: an efficient one-stage model for automatic lung tumor detection based on ct images, Mathematics 11~(10) (2023) 2344.

\bibitem{ega2023study}
A.~V. EGA, W.~ARDIATNA, Study on image processing method and data augmentation for chest x-ray nodule detection with yolov5 algorithm, ELKOMIKA: Jurnal Teknik Energi Elektrik, Teknik Telekomunikasi, \& Teknik Elektronika 11~(2) (2023) 424.

\bibitem{goel2024improving}
L.~Goel, P.~Patel, Improving yolov6 using advanced pso optimizer for weight selection in lung cancer detection and classification, Multimedia Tools and Applications (2024) 1--34.

\bibitem{mammeri2024early}
S.~Mammeri, M.~Amroune, M.-Y. Haouam, I.~Bendib, A.~Corr{\^e}a~Silva, Early detection and diagnosis of lung cancer using yolo v7, and transfer learning, Multimedia Tools and Applications 83~(10) (2024) 30965--30980.

\bibitem{csaman2023yolov8}
C.~{\c{S}}aman, {\c{S}}.~{\c{C}}elikba{\c{s}}, Yolov8-based lung nodule detection: A novel hybrid deep learning model proposal, International Research Journal of Engineering and Technology 10~(8) (2023) 230--237.

\bibitem{jain2023pulmonary}
S.~Jain, P.~Choudhari, M.~Gour, Pulmonary lung nodule detection from computed tomography images using two-stage convolutional neural network, The Computer Journal 66~(4) (2023) 785--795.

\bibitem{zhang2023lungseek}
H.~Zhang, H.~Zhang, Lungseek: 3d selective kernel residual network for pulmonary nodule diagnosis, The Visual Computer 39~(2) (2023) 679--692.

\bibitem{lu2024dual}
H.~Lu, K.~Liu, H.~Zhao, Y.~Wang, B.~Shi, Dual-layer detector spectral ct-based machine learning models in the differential diagnosis of solitary pulmonary nodules, Scientific Reports 14~(1) (2024) 4565.

\bibitem{tan2020efficientdet}
M.~Tan, R.~Pang, Q.~V. Le, Efficientdet: Scalable and efficient object detection, in: Proceedings of the IEEE/CVF conference on computer vision and pattern recognition, 2020, pp. 10781--10790.

\bibitem{zhou2019objects}
X.~Zhou, D.~Wang, P.~Kr{\"a}henb{\"u}hl, Objects as points, arXiv preprint arXiv:1904.07850 (2019).

\bibitem{liu2022stbi}
K.~Liu, Stbi-yolo: A real-time object detection method for lung nodule recognition, IEEE Access 10 (2022) 75385--75394.

\bibitem{ji2023lung}
Z.~Ji, Y.~Wu, X.~Zeng, Y.~An, L.~Zhao, Z.~Wang, I.~Ganchev, Lung nodule detection in medical images based on improved yolov5s, IEEE Access (2023).

\bibitem{hu2018squeeze}
J.~Hu, L.~Shen, G.~Sun, Squeeze-and-excitation networks, in: Proceedings of the IEEE conference on computer vision and pattern recognition, 2018, pp. 7132--7141.

\bibitem{woo2018cbam}
S.~Woo, J.~Park, J.-Y. Lee, I.~S. Kweon, Cbam: Convolutional block attention module, in: Proceedings of the European conference on computer vision (ECCV), 2018, pp. 3--19.

\bibitem{zhang2021sa}
Q.-L. Zhang, Y.-B. Yang, Sa-net: Shuffle attention for deep convolutional neural networks, in: ICASSP 2021-2021 IEEE International Conference on Acoustics, Speech and Signal Processing (ICASSP), IEEE, 2021, pp. 2235--2239.

\bibitem{liu2021nam}
Y.~Liu, Z.~Shao, Y.~Teng, N.~Hoffmann, Nam: Normalization-based attention module, arXiv preprint arXiv:2111.12419 (2021).

\bibitem{hou2021coordinate}
Q.~Hou, D.~Zhou, J.~Feng, Coordinate attention for efficient mobile network design, in: Proceedings of the IEEE/CVF conference on computer vision and pattern recognition, 2021, pp. 13713--13722.

\bibitem{wang2020eca}
Q.~Wang, B.~Wu, P.~Zhu, P.~Li, W.~Zuo, Q.~Hu, Eca-net: Efficient channel attention for deep convolutional neural networks, in: Proceedings of the IEEE/CVF conference on computer vision and pattern recognition, 2020, pp. 11534--11542.

\bibitem{wang2018non}
X.~Wang, R.~Girshick, A.~Gupta, K.~He, Non-local neural networks, in: Proceedings of the IEEE conference on computer vision and pattern recognition, 2018, pp. 7794--7803.

\bibitem{velivckovic2017graph}
P.~Veli{\v{c}}kovi{\'c}, G.~Cucurull, A.~Casanova, A.~Romero, P.~Lio, Y.~Bengio, Graph attention networks, arXiv preprint arXiv:1710.10903 (2017).

\bibitem{park2018bam}
J.~Park, S.~Woo, J.-Y. Lee, I.~S. Kweon, Bam: Bottleneck attention module, arXiv preprint arXiv:1807.06514 (2018).

\bibitem{huang2019ccnet}
Z.~Huang, X.~Wang, L.~Huang, C.~Huang, Y.~Wei, W.~Liu, Ccnet: Criss-cross attention for semantic segmentation, in: Proceedings of the IEEE/CVF international conference on computer vision, 2019, pp. 603--612.

\bibitem{dai2019second}
T.~Dai, J.~Cai, Y.~Zhang, S.-T. Xia, L.~Zhang, Second-order attention network for single image super-resolution, in: Proceedings of the IEEE/CVF conference on computer vision and pattern recognition, 2019, pp. 11065--11074.

\end{thebibliography}
